\documentclass[letterpaper, 10 pt, conference]{ieeeconf}  

\IEEEoverridecommandlockouts  

\let\labelindent\relax  

\usepackage{graphics} 
\usepackage{epsfig} 
\usepackage{mathptmx} 
\usepackage{times} 
\usepackage{amsmath} 
\usepackage{amssymb}  

\usepackage{comment}    
\usepackage{amssymb}    
\usepackage{mathtools}  
\usepackage[inline]{enumitem}   
\usepackage[svgnames]{xcolor}
\usepackage{bm}
\usepackage[caption=false]{subfig}
\usepackage{cite}

\usepackage{amsthm}     

\makeatletter
\let\NAT@parse\undefined
\makeatother
\usepackage[hidelinks]{hyperref}   

\DeclareMathAlphabet{\mathcal}{OMS}{cmsy}{m}{n} 

\theoremstyle{definition}

\makeatletter
\def\thm@space@setup{%
  \thm@preskip=\parskip \thm@postskip=\parskip
}
\makeatother

\newtheoremstyle{mythmstyle}
{}
{}
{\itshape}
{}
{\bfseries}
{:}
{.5em}
{{\bfseries\thmname{#1}\thmnumber{ #2}}\thmnote{ (#3)}}

\theoremstyle{mythmstyle}

\newtheorem{theorem}{Theorem}
\newtheorem{definition}{Definition}

\newtheorem{lemma}{Lemma}
\newtheorem{assumption}{Assumption}

\newtheorem{rem}{Remark}

\renewcommand{\vec}[1]{\mathbf{#1}}         
\newcommand{\T}{^{\mathsf{T}}}

\newcommand{\B}[1]{\if#1\relax\bm{#1}\else\mathbf{#1}\fi} 
\newcommand{\R}[1]{\mathrm{#1}}						      
\newcommand{\C}[1]{\mathcal{#1}}
\newcommand{\BB}[1]{\mathbb{#1}}
\newcommand{\unitvec}[1]{ \vec{\hat{#1}} }
\newcommand{\norm}[1]{\left\lVert #1 \right\rVert}
\newcommand{\abs}[1]{\left\lvert #1 \right\rvert}
\newcommand{\tr}[0]{\mathrm{trace}}
\newcommand{\remark}[1]{{\color{red} [#1]}}

\allowdisplaybreaks
\title{\LARGE \bf
Local convergence of 
multi-agent systems towards triangular patterns}%

\author{{Andrea Giusti$^{1}$, Marco Coraggio$^{2}$, and Mario di Bernardo$^{1, 2}$}
\thanks{This work was in part supported by the Research Project ``SHARESPACE'' funded by the European Union (EU HORIZON-CL4-2022-HUMAN-01-14. SHARESPACE. GA 101092889 - http://sharespace.eu),
and by the Research Project ``Centro Nazionale HPC, Big Data e Quantum Computing Italian Center for Super Computing (ICSC)'', funded by European Union (PNRR CN00000013).
}
\thanks{$^{1}$Department of Electrical Engineering and Information Technology, University of Naples Federico II, Via Claudio 21, Naples, 80125, Italy.}
\thanks{$^{2}$Scuola Superiore Meridionale, School for Advanced Studies, Largo S. Marcellino 10, Naples, 80138, Italy.}
\thanks{Contacts: 
        {\{andrea.giusti, marco.coraggio, mario.dibernardo\}@unina.it.}
}
}

\begin{document}

\maketitle
\thispagestyle{empty}
\pagestyle{empty}


\begin{abstract}
Geometric pattern formation is an important emergent behavior in many applications involving large-scale multi-agent systems, such as sensor
networks deployment and collective transportation.
Attraction/repulsion virtual forces are the most common control approach to achieve such behavior in a distributed and scalable manner.
Nevertheless, for most existing solutions only numerical and/or experimental evidence of their convergence is available.
Here, we revisit the problem of achieving pattern formation giving sufficient conditions to prove analytically that under the influence of appropriate virtual forces, a large-scale multi-agent swarming system locally converges towards a stable and robust triangular lattice configuration. 
%
%
Specifically, the proof is carried out using LaSalle's invariance principle and geometry-based arguments.
Our theoretical results are complemented by exhaustive numerical simulations confirming their effectiveness and estimating the region of asymptotic stability of the triangular configuration.
\end{abstract}

\section{Introduction}
\label{sec::intro}


Many natural and artificial systems consist of multiple interacting agents; their behavior being determined by both the individual agent dynamics and their interaction.
In some applications the number of \emph{agents} can be extremely large (\emph{large-scale multi-agent systems}) and the role played by their interconnections becomes predominant over their individual dynamics \cite{PShi2021}. Examples include cell populations \cite{Grandel2021}, swarming multi-robot systems \cite{Heinrich2022}, social networks \cite{Jusup2022} among many others.
%
%
Some of the most relevant emerging behavior exhibited by these systems involve their 
\emph{spatial organization, coordination}, and \emph{cooperation} \cite{Brambilla2013}.
%
A notable case is \emph{geometric pattern formation} \cite{HOh2017} where the agents are required to self-organize
into some desired \emph{pattern}, such as, for example, triangular lattices consisting of repeating adjacent triangles.
%
Applications of pattern formation include sensor networks deployment \cite{Zhao2019}, collective transportation and construction \cite{Rubenstein2013, Gardi2022}, and exploration and mapping \cite{Kegeleirs2021}.

%



Most of the existing distributed control algorithms for geometric pattern formation rely on the use of \emph{virtual forces} (or \emph{virtual potentials}), \cite{Giusti2022, Spears2004, Casteigts2012, Zhao2019, Torquato2009, Olfati-Saber2002IFAC, Mesbahi2010, Sakurama2021, Olfati-Saber2006}.
Within this framework, agents move under the effect of forces generated by the presence of their neighboring agents and the environment, causing attraction, repulsion, alignment, etc. 

Interestingly, most strategies are validated only numerically or experimentally
\cite{Giusti2022, Spears2004, Casteigts2012, Zhao2019}.
Among the exceptions, in \cite{Lee2008}, a geometric control approach based on trigonometric functions is proposed to build triangular lattices, and its  global convergence is proved. 
The extension to 3D spaces is validated analytically in \cite{Lee2010}.
Moreover, \emph{harmonic approximation} \cite{Hinsen2005} provides necessary conditions for the local stability of a lattice. 
These conditions are used in \cite{Torquato2009} to numerically design a virtual force that locally stabilizes an hexagonal lattice.
%
%
A general analysis of the effects of attraction/repulsion virtual forces is carried out in \cite{Gazi2002}, where the authors prove that the agents converge inside a bounded region, even though the specific equilibrium configuration is not characterized.
We wish to remark here that formation control \cite{Sakurama2021, Olfati-Saber2002IFAC,  Mesbahi2010}
differs from geometric pattern formation because of a typically smaller number of agents (order of tens) with, possibly, unique identifiers, numerous roles for the agents
and often some coordinated motion of the agents.
%
Similarly, when solving \emph{flocking} control problems, the emergence of coordinated motion is the crucial concern \cite{Olfati-Saber2006, FWang2022, GWang2022}.




In this paper, we revisit the problem of  geometric pattern formation using \emph{attraction/repulsion} virtual forces with the aim of bridging a gap in the existing literature and deriving a general proof of convergence when considering the formation of triangular lattice  configurations.
When compared to previous work, our stability results (i) can be applied to most control laws based on virtual forces (or potentials), rather than only holding for specific algorithms, e.g. \cite{Lee2008}, (ii) are sufficient rather than being only necessary \cite{Hinsen2005}, (iii) characterize the asymptotic configuration of the agents, rather than just proving its  boundedness \cite{Gazi2002}, and (iv) guarantee the emergence of triangular lattices rather than less regular ones, e.g. $\alpha$-lattices studied in \cite{Olfati-Saber2006}.

\section{Mathematical preliminaries}
\label{sec:preliminaries}
 
Given a vector $\vec{v} \in \BB{R}^d$, we denote by $[\vec{v}]_i$ its $i$-th element, by $\norm{\vec{v}}$ its Euclidean norm, and by $\unitvec{v}\coloneqq\frac{\vec{v}}{\norm{\vec{v}}}$ its direction. 
$\vec{0}$ denotes a column vector of appropriate dimension with all elements equal to 0.
Given a matrix $\vec{A}$, $[\vec{A}]_{ij}$ is its $(i, j)$-th element.

Given a continuous-time, autonomous dynamical system
\begin{equation}\label{eq:dynamical_system}
    \dot{\vec{x}}(t) = \vec{f}(\vec{x}(t)), \quad \vec{x}(0) = \vec{x}_0,
\end{equation}
 with state vector $\vec{x}(t) \in \BB{R}^d$, and $\vec{x}_0 \in \BB{R}^d$, we term as $\vec{\phi}(t, \vec{x}_0)$ its trajectory starting from $\vec{x}(0) = \vec{x}_0$.

\begin{definition}[Equilibrium set]
\label{def:equilibrium_set}
A set $\Xi  \subset \BB{R}^d$ is an \emph{equilibrium set} for system \eqref{eq:dynamical_system} if $ \vec{f}(\vec{x}) = \vec{0} \ \forall \vec{x} \in \Xi$.
\end{definition}

\begin{definition}[Local asymptotic stability {\cite[Definition~1.8]{Kuznetsov2004}}]
\label{def:LAS}
An equilibrium set $\Xi$ for system \eqref{eq:dynamical_system} is \emph{locally asymptotically stable} if $\forall \epsilon>0, \exists \delta > 0$ such that if $\min_{\vec{y} \in \Xi}\norm{\vec{x}_0 - \vec{y}} < \delta $, then
\begin{enumerate}
    \item 
    $\min_{\vec{y} \in \Xi}\norm{\vec{\phi}(t, \vec{x}_0)-\vec{y}} < \epsilon,  \ \forall t>0$, and
    \item
    $ \lim_{t \rightarrow +\infty} \vec{\phi}(t, \vec{x}_0) \in \Xi$.
\end{enumerate}
\end{definition}




\begin{definition}[Incidence matrix]
\label{def:incidence_mat}
Given a digraph with $n$ vertices and $m$ edges, its \emph{incidence matrix} $\vec{B}\in \mathbb{R}^{n\times m}$ has elements defined as
\begin{equation*}
    [\vec{B}]_{ij} \coloneqq \begin{dcases}
    + 1, &\text{if edge $j$ starts from vertex $i$},\\
    - 1, &\text{if edge $j$ ends in vertex $i$},\\
    0,   &\text{otherwise}.
    \end{dcases}
\end{equation*}
\end{definition}
%

\begin{definition}[Framework {\cite[p.~120]{Mesbahi2010}}]\label{def:framework}
Consider a (di\mbox{-)}graph $\C{G}=(\C{V}, \C{E})$ with $n$ vertices, and a set of positions $\vec{p}_1, \dots, \vec{p}_n \in \mathbb{R}^d$ associated to its vertices, with $\vec{p}_i \neq \vec{p}_j \ \forall i,j \in \{1, \dots, n\}$.
A \emph{$d$-dimensional framework} is the pair $(\C{G},\bar{\vec{p}})$, where $\bar{\vec{p}} \coloneqq [\vec{p}_1\T \ \cdots \ \vec{p}_n\T]\T \in \mathbb{R}^{dn}$.
Moreover, the \emph{length} of an edge, say $(i,j)\in \C{E}$, is $\norm{\vec{p}_i - \vec{p}_j}$.
\end{definition}

\begin{definition}[Congruent frameworks {\cite[p. 3]{Jackson2007}}]\label{def:congurent_framework}
Given a graph $\C{G}=(\C{V},\C{E})$ and two frameworks $(\C{G},\bar{\vec{p}})$ and $(\C{G},\bar{\vec{q}})$, these are \emph{congruent} if $\norm{\vec{p}_i - \vec{p}_j}=\norm{\vec{q}_i - \vec{q}_j}\ \forall i,j \in \C{V}$.
\end{definition}
%


\begin{definition}[Rigidity matrix {\cite[p.~5]{Jackson2007}}]
\label{def:rigidity_matrix}
Given a $d$-dimensional framework with $n\geq2$ vertices and $m$ edges, its \emph{rigidity matrix} $\vec{M}\in \mathbb{R}^{m\times dn}$ has elements defined as
\begin{equation}\label{eq:rigidity_matrix}
    [\vec{M}]_{e,(jd-d+k)}\coloneqq\begin{cases}
    [\vec{p}_j-\vec{p}_i]_k, & \parbox[t]{4cm}{ if edge $e$ starts from vertex $i$ and ends in vertex $j$,}\\
    [\vec{p}_i-\vec{p}_j]_k, & \parbox[t]{4cm}{ if edge $e$ starts from vertex $j$ and ends in vertex $i$,}\\
    0, &\mbox{otherwise.}
    \end{cases}
\end{equation}
with $k=1, \dots, d$.
\end{definition}

\begin{definition}[Infinitesimal rigidity {\cite[p.~122]{Mesbahi2010}}]
\label{def:inf_rigidity}
A framework with rigidity matrix $\vec{M}$ is \emph{infinitesimally rigid} if, for any infinitesimal motion, say $\vec{u}$,%
\footnote{$\vec{u}$ can be interpreted as either a velocity or a small displacement.}
of its vertices, such that the length of the edges is preserved, it holds that $\vec{M u} = 0$.
\end{definition}

To give a geometrical intuition of the concept of infinitesimal rigidity, we note that an infinitesimally rigid framework is also rigid \cite[p.~122]{Mesbahi2010}, according to the definition below.%
\footnote{In rare cases, a rigid graph is not infinitesimally rigid; e.g. \cite[p.~7]{Jackson2007}.}

\begin{definition}[Rigidity {\cite[p. 3]{Jackson2007}}]
\label{def:rigidity}
A framework is \emph{rigid} if every continuous motion of the vertices, that preserves the length of the edges, also preserves the distances between all pairs of vertices.
\end{definition}
As a consequence, in a rigid framework, any continuous motion that does \emph{not} preserve the distance between any two pairs of vertices also does \emph{not} preserve the length of at least one edge.

\begin{theorem}[{\cite[p. 122]{Mesbahi2010}}]
\label{th:rigidity}
A 2-dimensional framework with $n\geq2$ vertices and rigidity matrix $\vec{M}$ is infinitesimally rigid if and only if $\R{rank}(\vec{M})=2n-3$.

\end{theorem}



\begin{definition}[Swarm]
\label{def:swarm}
A \emph{(planar) swarm} $\C{S} \coloneqq \{1,2,\dots,n\}$ is a set of $n \in \mathbb{N}_{>0}$ identical agents that can move on the plane.
For each agent $i \in \C{S}$, $\vec{x}_i(t)\in \mathbb{R}^2$ denotes its position in the plane at time $t \in \mathbb{R}_{\geq0}$.
\end{definition}

Moreover, we call $\bar{\vec{x}}(t) \coloneqq [\vec{x}_1\T(t) \ \cdots \ \vec{x}_n\T(t)]\T \in \mathbb{R}^{2n}$ the \emph{configuration} of the swarm, define $\vec{x}_{\R{c}}(t) \coloneqq \frac{1}{n} \sum_{i = 1}^n \vec{x}_i(t) \, \in \mathbb{R}^2$ as its \emph{center}, and
denote by $\vec{r}_{ij}(t) \coloneqq \vec{x}_{i}(t)-\vec{x}_{j}(t) \in \mathbb{R}^2$ the relative position of agent $i$ with respect to agent $j$.



\begin{definition}[Adjacency set]\label{def:adjacency_set}
Given a swarm $\C{S}$, the \emph{adjacency set} of agent $i$ at time $t$ is 
$
    \C{A}_i(t) \coloneqq \{ j \in \C{S} \setminus \{i\} : \Vert \vec{r}_{ij}(t)\Vert \leq R_\mathrm{a} \}
$,
where $R_\R{a} \in \BB{R}_{>0}$ is the \emph{maximum link length}.
\end{definition}

\begin{definition}[Links]
\label{def:links}
A \emph{link} is a pair $(i,j) \in \C{S} \times \C{S}$ such that $j \in \C{A}_i(t)$; $\norm{\vec{r}_{ij}(t)}$ is its \emph{length}.
The set of all links existing in a certain configuration $\bar{\vec{x}}$ is denoted by $\C{E}(\bar{\vec{x}})$.
\end{definition}
Notice that $(i,j) \in \C{E}(\bar{\vec{x}}) \iff (j,i) \in \C{E}(\bar{\vec{x}})$.



\begin{definition}[Swarm graph and framework]
\label{def:swram_graph}
The \emph{swarm graph} is the digraph $\C{G}(\bar{\vec{x}})\coloneqq(\C{S},\C{E}(\bar{\vec{x}}))$.
The \emph{swarm framework} is $\C{F}(\bar{\vec{x}})\coloneqq(\C{G}(\bar{\vec{x}}), \bar{\vec{x}})$.
\end{definition}

\begin{figure}[t]
    \centering
    \subfloat[]{
    \includegraphics[width=0.35\columnwidth]{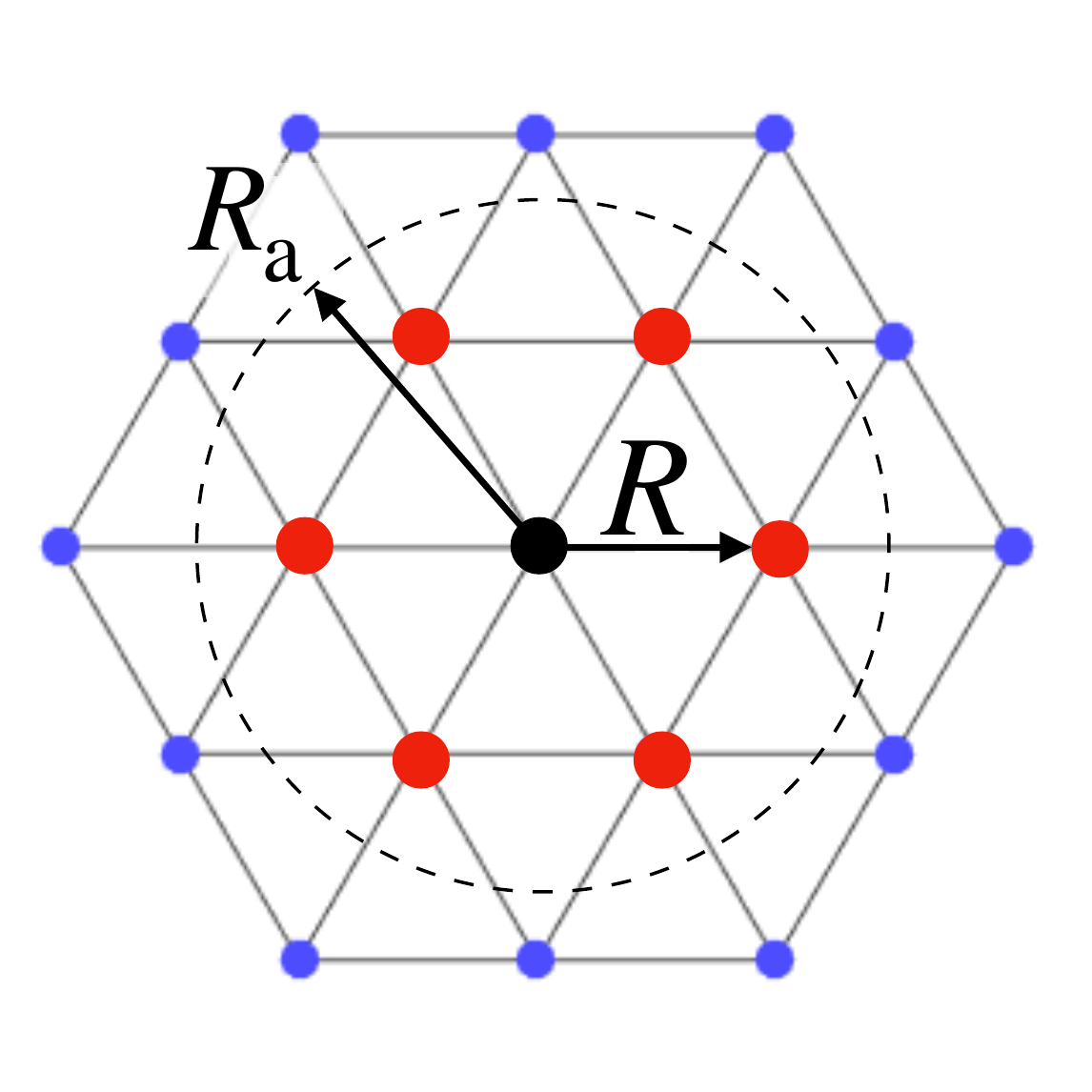}\label{fig:R_a_R}} \qquad
    \subfloat[]{
    \includegraphics[trim=4mm 4mm 4mm 4mm, clip, width=0.35\columnwidth]{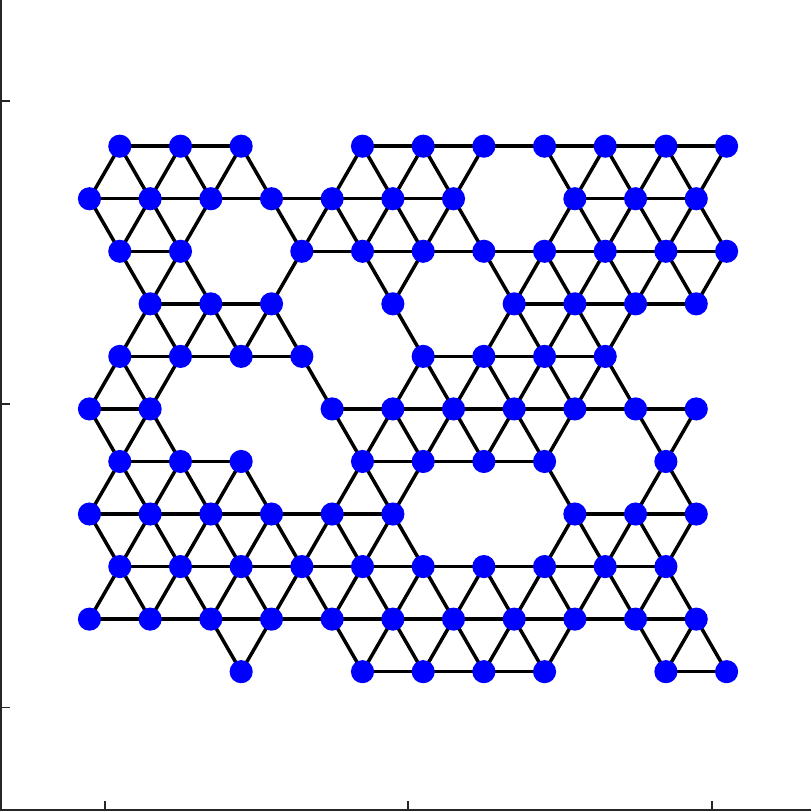}}
    
    \caption{
    Triangular configurations. 
    (a) Schematic representation of a triangular lattice; red agents belong to the adjacency set of the black agent.
    (b) Example of a triangular configuration with $n=100$ agents.}
    \label{fig:triangular_lattices_examples}
\end{figure}

\begin{definition}[Triangular lattice configuration]
\label{def:triangular_lattice}
Consider a planar swarm $\C{S}$ with framework $\C{F}(\bar{\vec{x}}^*)$.
$\bar{\vec{x}}^*$ 
 is a \emph{triangular (lattice) configuration} if
\begin{enumerate}[label=(\Alph*)]
    \item\label{condition:rigidity}
    $\C{F}(\bar{\vec{x}}^*)$
    is infinitesimally rigid, and
    \item\label{condition:link_length} 
    $\norm{\vec{r}_{ij}}=R,\ \forall (i,j)\in \C{E}(\bar{\vec{x}}^*)$,
\end{enumerate}
where $R \in \BB{R}_{>0}$ denotes the \emph{desired link length}.
\end{definition}
Here, we assume that
\begin{equation}\label{eq:R_a}
    R_\R{a}\in ]R, R\sqrt{3}[,
\end{equation}
so that, when the swarm is in a triangular configuration, the adjacency set (Definition~\ref{def:adjacency_set}) of any agent includes only the agents in its immediate surroundings, and all the links (Definition~\ref{def:links}) have length $R$ (see Fig. \ref{fig:triangular_lattices_examples}).

We denote by $\C{T} \subset \BB{R}^{2n}$ the set of all triangular lattice configurations; it is immediate to verify that $\C{T}$ is unbounded and disconnected.  

\begin{figure}[t]
    \centering
    \subfloat[]{
    \includegraphics[width=0.45\columnwidth]{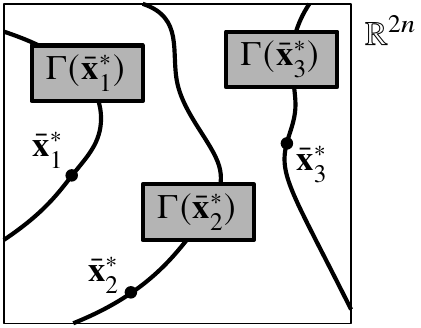}
    \label{fig:triangular_lattices}}
    \subfloat[]{
    \includegraphics[width=0.45\columnwidth]{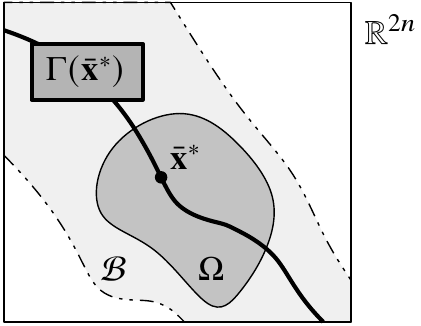}
    \label{fig:all_sets}}
    \caption{(a): Sets of triangular lattices configurations. (b): Sets used in the proof of Theorem \ref{th:local_stability_lyap}.}
\end{figure}

\begin{definition}[Congruent configurations]\label{def:congruent_conf}
Given a configuration $\bar{\vec{x}}^\diamond$, we define \emph{the set of its congruent configurations} $\Gamma(\bar{\vec{x}}^\diamond)$ as the set of configurations with congruent associated frameworks (see Definition~\ref{def:congurent_framework}), that is
\begin{equation*}
    \Gamma(\bar{\vec{x}}^\diamond) \coloneqq \{ \bar{\vec{x}} \in \BB{R}^{2n} : \norm{\vec{x}_{i}-\vec{x}_{j}} = \norm{\vec{x}_{i}^\diamond-\vec{x}_{j}^\diamond}, \forall i,j \in \C{S} \}.
\end{equation*}
\end{definition}
These configurations are obtained by translations and rotations of the framework $\C{F}(\bar{\vec{x}}^\diamond)$; thus, it is immediate to verify that $\Gamma(\bar{\vec{x}}^\diamond)$ is connected and unbounded for any $\bar{\vec{x}}^\diamond$ (see Fig. \ref{fig:triangular_lattices}).
Also, note that $\bar{\vec{x}}^* \in \C{T} \iff \Gamma(\bar{\vec{x}}^*) \subset \C{T}$, and
\begin{equation}
    \label{eq:TasUnion}
    \C{T}=\bigcup_{\bar{\vec{x}}^* \in \C{T}} \Gamma(\bar{\vec{x}}^*).
\end{equation}

In the following, we 
omit the dependence on time when clear from the context.

\section{Problem Statement}
Consider a swarm $\C{S}$ of $n$ agents, with agents' dynamics described by
\begin{equation}\label{eq:model}
    \dot{\vec{x}}_i(t) = \vec{u}_i(t), \ \ \forall i \in \C{S},
\end{equation}
where 
$\vec{u}_i(t)\in \mathbb{R}^2$ is a control law to be designed.

Let $R_\R{s} \in \BB{R}_{>0}$ be a \emph{sensing radius} and define the \emph{interaction set} of agent $i$ at time $t$ as 
\begin{equation}
    \C{I}_i(t) \coloneqq \{ j \in \C{S} \setminus \{i\} : \Vert \vec{r}_{ij}(t)\Vert \leq R_{\mathrm{s}} \}.
\end{equation}
For the term $\vec{u}_i(t)$ in \eqref{eq:model}, we consider the distributed \emph{virtual forces} control law, given by
\begin{equation}\label{eq:control_law}
    \vec{u}_i(t) \coloneqq \sum_{j \in \C{I}_i(t)} f\left( \norm{\vec{r}_{ij}(t)} \right)\, \unitvec{r}_{ij}(t),
\end{equation}
where $f : \mathbb{R}_{\geq 0} \rightarrow \mathbb{R}$ is the \emph{interaction function}.

Note that in general there is no specific relation between $\C{I}_i$ and $\C{A}_i$ (see Definition~\ref{def:adjacency_set}); however, we reasonably assume that $R_{\R{s}}\geq R_\R{a}$, so that
\begin{equation}\label{eq:A_subset_I}
    \C{A}_i \subseteq \C{I}_i, \quad \forall i\in \C{S}.
\end{equation}


The following result slightly extends the one reported in {\cite[Lemma 1]{Gazi2002}}.
\begin{lemma}\label{th:invariant_center}
The position of the center of the swarm (see Definition~\ref{def:swarm}), say $\vec{x}_{\R{c}}$, under the control law \eqref{eq:control_law} is invariant, that is 
$
    \dot{\vec{x}}_{\R{c}} = \vec{0}\ \forall \bar{\vec{x}} \in \BB{R}^{2n}.
$
\end{lemma}
\begin{proof}
Exploiting \eqref{eq:model} and \eqref{eq:control_law}, the dynamics of the center of the swarm is given by
\begin{equation}
    \label{eq:velocity_cm}
    \dot{\vec{x}}_{\R{c}} \coloneqq \frac{1}{n} \sum_{i = 1}^n \dot{\vec{x}}_i = \frac{1}{n} \sum_{i = 1}^n {\vec{u}}_i =
    \frac{1}{n} \sum_{i = 1}^n \sum_{j \in \C{I}_i} f(\norm{\vec{r}_{ij}})\, \unitvec{r}_{ij}.
\end{equation}
Since in a swarm the existence of any link $(i,j)$ implies the existence of link $(j,i)$ (see Definition~\ref{def:adjacency_set}), in \eqref{eq:velocity_cm}, for any term $f(\norm{\vec{r}_{ij}}) \, \unitvec{r}_{ij}$ there exists a term $f(\norm{\vec{r}_{ji}}) \, \unitvec{r}_{ji}=-f(\norm{\vec{r}_{ij}}) \, \unitvec{r}_{ij}$ (because $\norm{\vec{r}_{ij}} = \norm{\vec{r}_{ji}}$ and $\unitvec{r}_{ij} = - \unitvec{r}_{ji}$).
Therefore, the sum of the two is zero, yielding the thesis.
\end{proof}

\section{Convergence to a triangular configuration}
\label{sec:main_results}
We can now state the main result of this work, showing that, given an interaction function $f$ (in \eqref{eq:control_law}) that generates short range repulsion and long range attraction, the set of triangular configurations of the swarm is a locally asymptotically stable equilibrium set (see Definitions \ref{def:equilibrium_set} and \ref{def:LAS}).

\begin{assumption}\label{ass:interaction_function}
    $f$ (in \eqref{eq:control_law}) is such that:
\begin{enumerate}[label=(a\arabic*)]
    \item \label{hp:null_point} $f(R)=0$,
    \item \label{hp:attraction_repulsion} $f(z) > 0$ for $z \in [0;R [$ and $f(z) < 0$ for $z>R$, 
    \item \label{hp:integrable} $f(z)$ is continuous in $[0; R_\R{a}]$,
    \item \label{hp:vanishing} $f(z) = 0$ for any $z > R_\R{a}$,
    \end{enumerate}
\end{assumption}

\begin{figure}[t]
    \centering
    \includegraphics{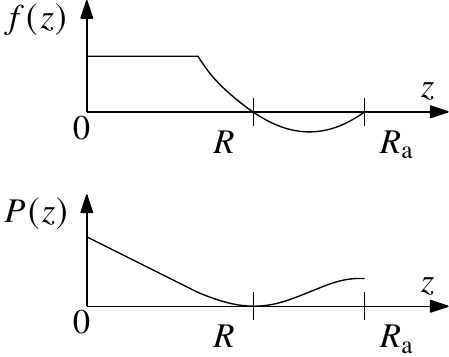}
    \caption{Example of an interaction function $f$ (top panel) and its corresponding potential $P$ (bottom panel).}
    \label{fig:interaction_function_and_potential}
\end{figure}

An exemplary interaction function fulfilling the assumption above is portrayed in Fig. \ref{fig:interaction_function_and_potential}.

Without loss of generality, we further assume that, under Assumption \ref{ass:interaction_function}, in a sufficiently small neighborhood of a triangular configuration, all other equilibria are also triangular (supporting evidence showing that this assumption is not restrictive is reported in the \hyperref[sec:appendix]{Appendix}).

\begin{theorem}\label{th:local_stability_lyap}
Let Assumption \ref{ass:interaction_function} hold.
Then, for any triangular configuration $\bar{\vec{x}}^*$, 
$\Gamma(\bar{\vec{x}}^*)$ is a locally asymptotically stable equilibrium set.
Consequently, $\C{T}$ is also a locally asymptotically stable equilibrium set.
\end{theorem}
\begin{proof}
Let us consider \emph{any} triangular configuration $\bar{\vec{x}}^* \in \C{T}$, 
with center $\vec{x}^*_{\R{c}} \coloneqq \frac{1}{n} \sum_{i = 1}^n \vec{x}_i^*$ and relative positions $\B{r}_{ij}^*$,
and the set $\Gamma(\bar{\vec{x}}^*)$ of its congruent configurations.
Recalling Definition~\ref{def:triangular_lattice}.\ref{condition:link_length} and \ref{hp:null_point}, we have that $\bar{\vec{x}}^*$ is an equilibrium point of \eqref{eq:model}--\eqref{eq:control_law}; thus, $\Gamma(\bar{\vec{x}}^*)$ and $\C{T}$ are equilibrium sets.
Next, we will prove local asymptotic stability
of $\Gamma(\bar{\vec{x}}^*)\subset \C{T}$, which implies local asymptotic stability of $\C{T}$ through \eqref{eq:TasUnion}.

\paragraph*{Step 1 (Lyapunov function)}

Given a configuration $\bar{\vec{x}} \in \BB{R}^{2n}$ with center $\vec{x}_{\R{c}}$ and inducing the links in $\C{E}(\bar{\vec{x}})$ according to Definition~\ref{def:links}, let $m \coloneqq \lvert \C{E}(\bar{\vec{x}}) \rvert$ and order the links in $\C{E}(\bar{\vec{x}})$ arbitrarily, so that $\vec{r}_1, \dots, \vec{r}_m$ refer to the relative positions $\vec{r}_{ij}$ for $(i,j)\in \C{E}(\bar{\vec{x}})$.
Recalling \ref{hp:integrable}, we can define the potential function
$P:[0, R_\R{a}]\to \BB{R}$ given by $P(z)=- \int_{R}^{z} f(y) \, \R{d}y$ (see Fig. \ref{fig:interaction_function_and_potential}). 
Note that $P(R)=0$, $\frac{\R{d}P}{\R{d} z}(z)= -f(z)$, and, from \ref{hp:attraction_repulsion},
\begin{equation} \label{eq:P_positive}
    P(z) > 0 \quad \forall z \in \BB{R}_{\geq0} \setminus \{R\} .    
\end{equation}

Then, let us consider the candidate Lyapunov function
\begin{equation}\label{eq:V+cm}
\begin{aligned}
    V(\bar{\vec{x}}) &\coloneqq \norm{\vec{x}^*_\R{c} - \vec{x}_\R{c} }^2 + \sum_{k\in\C{E}(\bar{\vec{x}})} P(\norm{\vec{r}_{k}}).
\end{aligned}
\end{equation}
By \eqref{eq:P_positive}, it holds that
    $V(\bar{\vec{x}})\geq0 \ \forall \bar{\vec{x}} \in \BB{R}^{2n}$,
and $V = 0$ if and only if both $\vec{x}_{\R{c}} =\vec{x}_{\R{c}}^*$ and Definition~\ref{def:triangular_lattice}.\ref{condition:link_length} holds%
. 

\paragraph*{Step 2 (Properties of $V$)}

$V(\bar{\vec{x}})$ is discontinuous over $\BB{R}^{2n}$ (because $\C{E}(\bar{\vec{x}})$ changes when links (dis-)appear).
However, $V(\bar{\vec{x}})$ is continuous and differentiable in any subset of $\BB{R}^{2n}$ where the set $\C{E}(\bar{\vec{x}})$ of links is constant.
To find such a set, we seek conditions on $\vec{\bar{x}}$ such that $\C{E}(\vec{\bar{x}}) = \C{E}(\vec{\bar{x}}^*)$ (see Definitions \ref{def:adjacency_set} and \ref{def:links}), i.e.,
\begin{subequations}
\begin{align}
    \norm{\vec{r}_{ij}} &< R_\R{a}, \quad \forall (i, j) \in \C{E}(\vec{\bar{x}}^*),\label{eq:links_preserved}\\
    \norm{\vec{r}_{ij}} &> R_\R{a}, \quad \forall (i, j) \not\in \C{E}(\vec{\bar{x}}^*).\label{eq:no_links_created}
\end{align}
\end{subequations}
\eqref{eq:links_preserved} means that all links in $\C{E}(\vec{\bar{x}}^*)$ are preserved in $\C{E}(\vec{\bar{x}})$, while \eqref{eq:no_links_created} means that no new links are created in $\C{E}(\vec{\bar{x}})$ with respect to $\C{E}(\vec{\bar{x}}^*)$.
With simple algebraic manipulations it is possible to show that  \eqref{eq:links_preserved} and \eqref{eq:no_links_created} hold if $\bar{\vec{x}} \in \C{B}$, where
\begin{equation}
    \label{eq:setB}
    \C{B} \coloneqq \{ \bar{\vec{x}} \in \mathbb{R}^{2n} : \abs{ \norm{\vec{r}_{ij}} - \norm{\vec{r}_{ij}^*} } < \beta, \ \forall i,j \in \C{S} \},
\end{equation}
and $\beta < \min_{i,j\in \C{S}}  \abs{ R_\R{a} - \norm{\vec{r}_{ij}^*} }$.
Note that $\C{B}$ can be interpreted as a ``neighborhood'' of $\Gamma(\bar{\vec{x}}^*)$ with ``width'' $\beta$ (see Fig. \ref{fig:all_sets}). 
Hence, $\C{E}(\vec{\bar{x}}) = \C{E}(\vec{\bar{x}}^*)$ in $\C{B}$, and thus $V$ is continuously differentiable in $\C{B}$.

%


\paragraph*{Step 3 (Analysis of $\dot{V}$)}

At this point, we can restrict our analysis to the set $\C{B}$ to study the attractivity of $\Gamma(\bar{\vec{x}}^*)$.
Let us start by studying the dynamics of the agents.
From \eqref{eq:model}--\eqref{eq:control_law}, we have
\begin{equation}\label{eq:x_i_dot}
    \dot{\vec{x}}_{i}= \sum_{j\in \C{I}_i} f(\Vert \vec{r}_{ij}\Vert)  \unitvec{r}_{ij}.
\end{equation}
Hypothesis \ref{hp:vanishing} and \eqref{eq:A_subset_I} imply that in \eqref{eq:x_i_dot} we have
\begin{equation}\label{eq:replace_I_with_A}
    \sum_{\substack{j\in \C{I}_i}} f(\norm{\vec{r}_{ij}})  \unitvec{r}_{ij} =
    \sum_{\substack{j\in \C{A}_i}} f(\norm{\vec{r}_{ij}})  \unitvec{r}_{ij}.
\end{equation}
Then, exploiting \eqref{eq:replace_I_with_A} and the incidence matrix $\vec{B}$ (Definition~\ref{def:incidence_mat}) of the swarm graph,
\eqref{eq:x_i_dot} can be rewritten as%
\begin{equation} \label{eq:x_i_dot_Incidence}
\begin{aligned} 
    \dot{\vec{x}}_{i}= \sum_{j\in \C{A}_i} f(\Vert \vec{r}_{ij}\Vert)  \unitvec{r}_{ij} = \sum_{k=1}^m [\vec{B}]_{ik} f(\Vert \vec{r}_{k}\Vert) \unitvec{r}_{k}.
\end{aligned}
\end{equation}
Moreover we can write the dynamics of the relative positions along a link $k$ as
$
    \dot{\vec{r}}_{k}= \sum_{i=1}^n [\vec{B}]_{ik} \dot{\vec{x}}_{i}.
$ 
Thus, exploiting Lemma \ref{th:invariant_center} and \eqref{eq:x_i_dot_Incidence},
we get
\begin{equation*}
\begin{aligned}
    \dot{V}(&\bar{\vec{x}}) = \sum_{k=1}^m \frac{\partial V}{\partial \norm{\vec{r}_k}}\ \frac{\partial \norm{\vec{r}_k}}{\partial \vec{r}_k} \ \dot{\vec{r}}_k 
    = \sum_{k=1}^m P'(\norm{\vec{r}_k}) \ \unitvec{r}_k\T  \sum_{i=1}^n [\vec{B}]_{ik} \dot{\vec{x}}_{i} \\
    &=-\sum_{i=1}^n  \sum_{k=1}^m [\vec{B}\T]_{ki}  \ f(\norm{\vec{r}_k}) \ \unitvec{r}_k\T \dot{\vec{x}}_{i} 
    =-\sum_{i=1}^n \dot{\vec{x}}_i\T \dot{\vec{x}}_i = - \dot{\bar{\vec{x}}}\T \dot{\bar{\vec{x}}}.
\end{aligned}
\end{equation*}
We can hence conclude that $\dot{V}(\bar{\vec{x}})=0$ if and only if $\dot{\bar{\vec{x}}}=\vec{0}$, i.e., in correspondence of equilibrium configurations.

Now, choosing $\beta$ small enough, we can exclude the presence of equilibrium configurations not belonging to $\Gamma(\bar{\vec{x}}^*)$, and therefore 
\begin{equation}
\begin{cases}\label{eq:characterization_V_dot}
    \dot{V}(\bar{\vec{x}})=0, 
    &\mbox{if }\bar{\vec{x}} \in \Gamma(\bar{\vec{x}}^*),\\
    \dot{V}(\bar{\vec{x}}) < 0,
    &\mbox{if }\bar{\vec{x}} \in \C{B} \setminus \Gamma(\bar{\vec{x}}^*).
\end{cases}
\end{equation}


\paragraph*{Step 4 (Applying LaSalle's invariance principle)}
To complete the proof, we define a forward invariant neighborhood of $\bar{\vec{x}}^*$ and then apply LaSalle's invariance principle. 
Given some $\omega \in \BB{R}_{>0}$, let $\Omega$ be the largest connected set containing $\bar{\vec{x}}^*$ such that
$ V(\bar{\vec{x}}) \leq \omega \ \forall \bar{\vec{x}} \in \Omega$ (see Fig. \ref{fig:all_sets}).
In particular, we select $\omega$ small enough that $\Omega \subseteq \C{B}$.%
\footnote{Such a value of $\omega$ exists because $\C{B}$ is a ``neighborhood'' of $\Gamma(\bar{\vec{x}}^*)$ (in the sense of \eqref{eq:setB}) and,  by the rigidity of framework $\C{F}(\vec{\bar{x}}^*)$ (Definition~\ref{def:rigidity}), any continuous motion of the vertices that changes the distance between any two vertices also changes the length of at least one link, causing $V$ to increase.}
Since $V(\bar{\vec{x}}) \leq \omega$ and $\dot{V}(\bar{\vec{x}}) \leq 0$ for all $ \bar{\vec{x}} \in \Omega$, then $\Omega$ is forward invariant.
Moreover, 
$\Omega$ is closed, because $V$ is continuous in $\Omega$, and $\Omega$ is the inverse image of the closed set $[0,\omega]$.
$\Omega$ is also bounded because (i) translations too far from $\bar{\vec{x}}^*$ cause $V$ to increase beyond $\omega$ (see \eqref{eq:V+cm}), and (ii) $\Omega \subseteq \C{B}$ implies that the deformations of the framework are bounded (see \eqref{eq:setB}).
Since $\Omega$ is closed and bounded, it is also compact.

As $\Omega$ is compact and forward invariant, we can apply LaSalle's invariance principle \cite[Theorem~4.4]{Khalil2002}, and noting that, in $\Omega$, $\dot{V}(\bar{\vec{x}})=0$ if and only if $\bar{\vec{x}} \in \Gamma(\bar{\vec{x}}^*)$ (see \eqref{eq:characterization_V_dot}), we get that all the trajectories starting in $\Omega$ converge to $\Gamma(\bar{\vec{x}}^*) \cap \Omega$.
This and the forward invariance of $\Omega$ imply that $\Gamma(\bar{\vec{x}}^*)$ is locally asymptotically stable,
and so is $\C{T}$ because of \eqref{eq:TasUnion}.
\end{proof}


\section{Numerical validation}
\label{sec:validation}

In this section, we validate numerically the result presented in Section \ref{sec:main_results} and estimate the basin of attraction of $\C{T}$.

\subsection{Simulation setup}

We set the desired link length to $R=1$, the maximum link length to $R_\R{a}=  (1+\sqrt{3})/2 \approx 1.37$, the sensing radius to $R_\R{s}=3$, and the number of agents to $n=100$.

\begin{figure}[t]
    \centering
    \includegraphics[width=0.9\columnwidth]{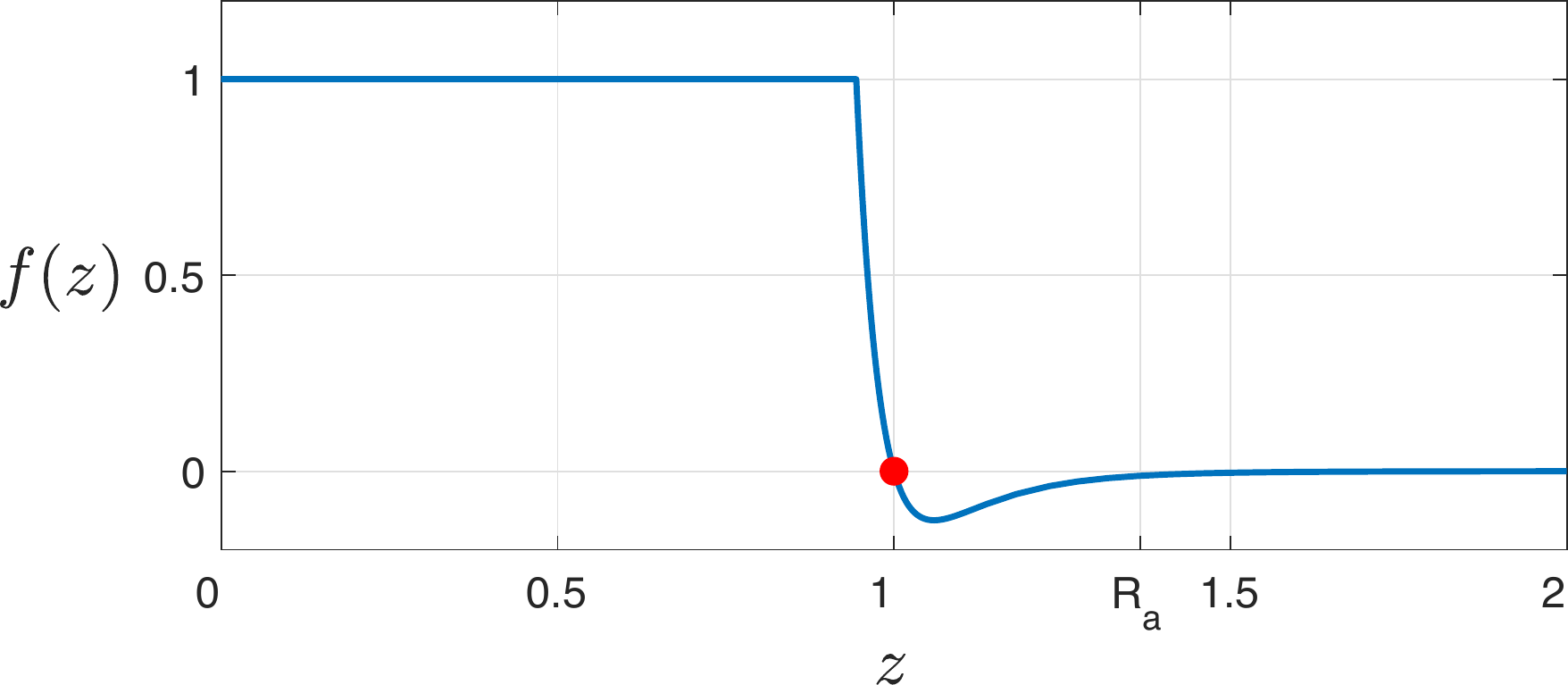}
    \caption{Plot of the interaction function defined by \eqref{eq::Lennard-Jones}. The zero of the function is highlighted by a red dot.}
    \label{fig:interaction_function}
\end{figure}

The interaction function $f$ is chosen as the Physics-inspired Lennard-Jones function \cite{Brambilla2013, Giusti2022},
given by
\begin{equation}\label{eq::Lennard-Jones}
    f(z) = \min \left\lbrace \left( \frac{a}{z^{2c}}-\frac{b}{z^c}\right), \ 1 \right\rbrace,
\end{equation}
where we select $a = b = 0.5$ and $c=12$; see Fig.~\ref{fig:interaction_function}.
In \eqref{eq::Lennard-Jones}, $f$ is saturated to $1$ to avoid divergence of $f$ for $z \to 0$.
Concerning Assumption \ref{ass:interaction_function}, the interaction function $f$ satisfies \ref{hp:null_point}, \ref{hp:attraction_repulsion}, and \ref{hp:integrable}.
Also, as shown in Fig.~\ref{fig:interaction_function}, it quickly tends to zero so that we can assume it practically satisfies \ref{hp:vanishing}.
The choice of not setting $f(z)$ exactly equal to zero for $z\geq R_a$ is intentional as it allows to account for long range attraction between the agents, which is frequently required in swarm robotics applications \cite{Gazi2002}.

To assess if the swarm is in a triangular configuration, we check the conditions in Definition~\ref{def:triangular_lattice}.
%
To evaluate whether a configuration is infinitesimally rigid, we use Theorem \ref{th:rigidity}.
Moreover, we define the \emph{error} 
$e(t) \coloneqq \max_{k\in\mathcal{E}(t)} \abs{\norm{\vec{r}_{k}(t)} - R}$,
which is zero when the configuration is triangular. 
Also, as long as $e(t)$ is lower than $R_\R{a}-R$, links in the configuration of interest are neither created nor destroyed.

For each simulation, the initial positions of the agents are obtained by picking a random triangular configuration and then applying, to each agent, a random displacement drawn from a uniform distribution over a disk of radius $\delta \in \BB{R}_{\ge 0}$.


All simulation are run in {\sc Matlab}%
\footnote{Code available at {\url{https://github.com/diBernardoGroup/SwarmSimPublic}}.} and last $20\, \text{s}$;
the agents' dynamics \eqref{eq:model}--\eqref{eq:control_law} are integrated using the forward Euler method with a fixed time step equal to $0.01\, \text{s}$.

\begin{rem}
\label{rem:extension_3D}
Theorem~\ref{th:local_stability_lyap}, together with Definitions~\ref{def:swarm} and \ref{def:triangular_lattice} allow a straightforward extension of the analysis to the three-dimensional case ($d = 3$). The only cumbersome step is to assess the infinitesimal rigidity of the 3D framework of interest as  Theorem~\ref{th:rigidity} can no longer be applied.
\end{rem}

\subsection{Numerical results}

\begin{figure}[t]
    \centering
    \subfloat[Terminal values of the metrics]{\includegraphics[width=1\columnwidth]{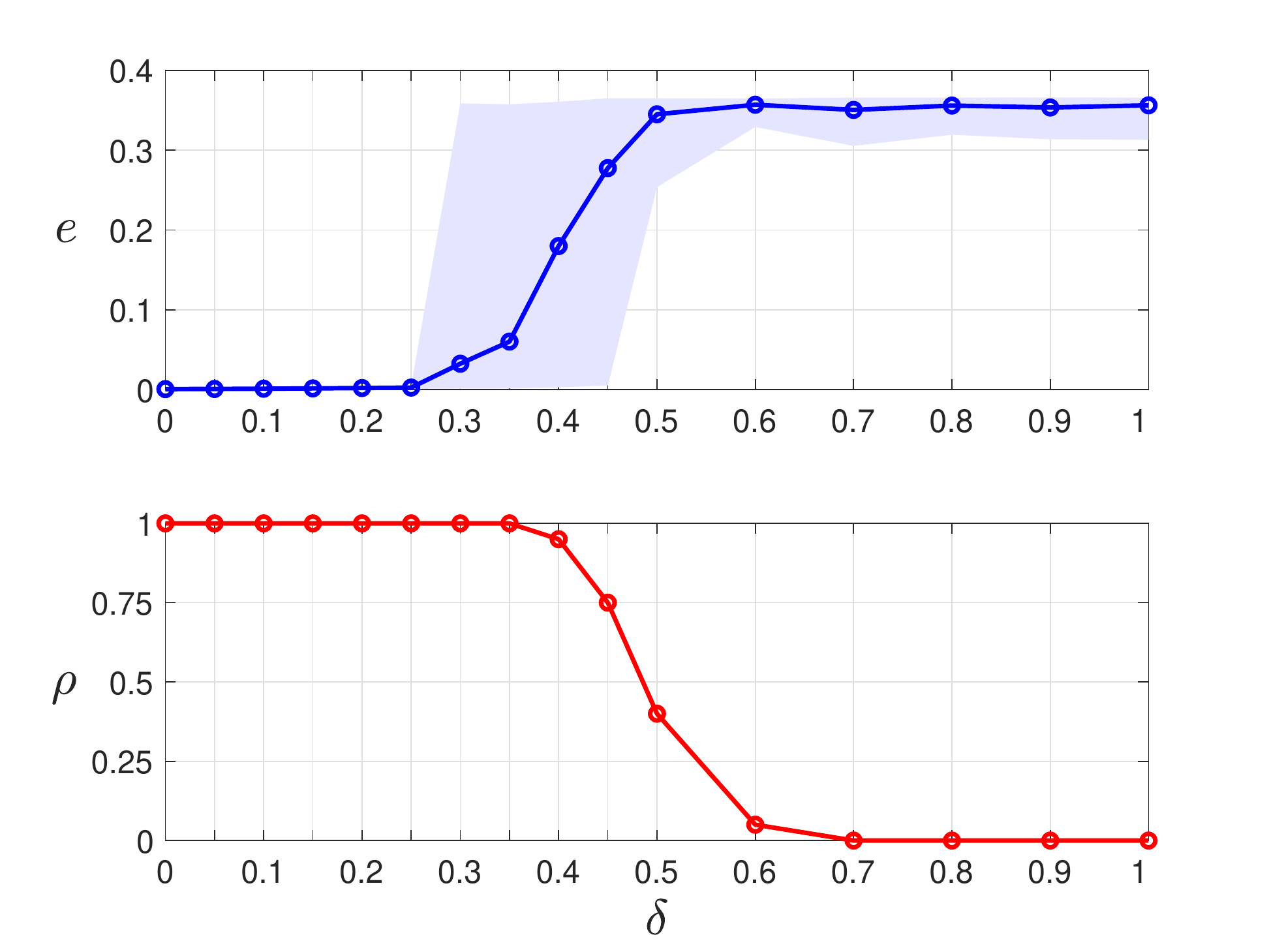}}\\[-0.5ex]
    \subfloat[Initial configurations]{
    \captionsetup[subfigure]{labelformat=empty}
    \subfloat[$\delta=0.2$]{\includegraphics[trim={2mm 2mm 2mm 12mm},clip,width=0.32\columnwidth]{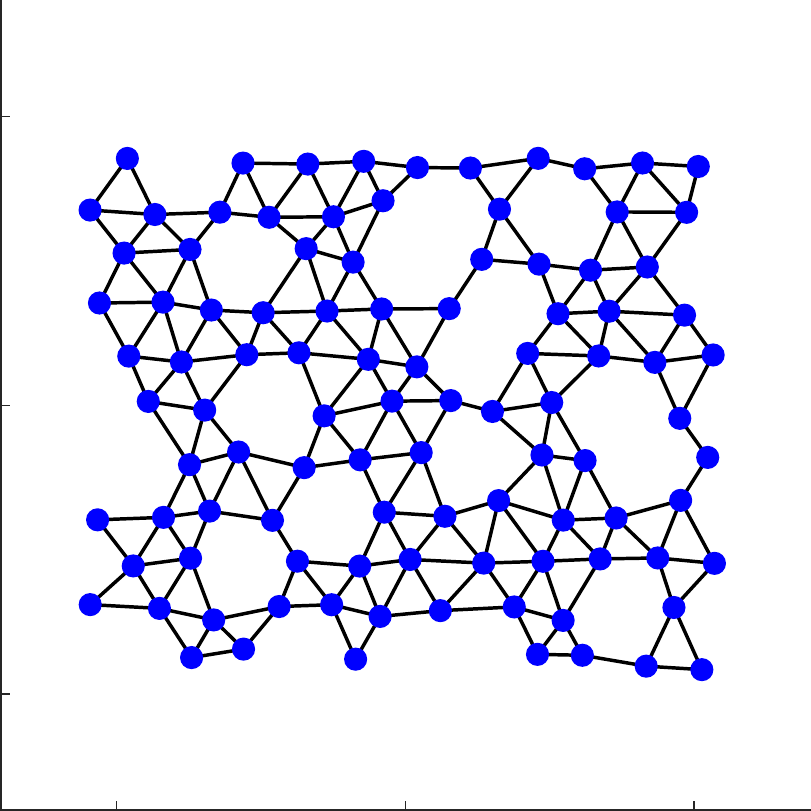}}
    \subfloat[$\delta=0.4$]{\includegraphics[trim={2mm 2mm 2mm 12mm},clip,width=0.32\columnwidth]{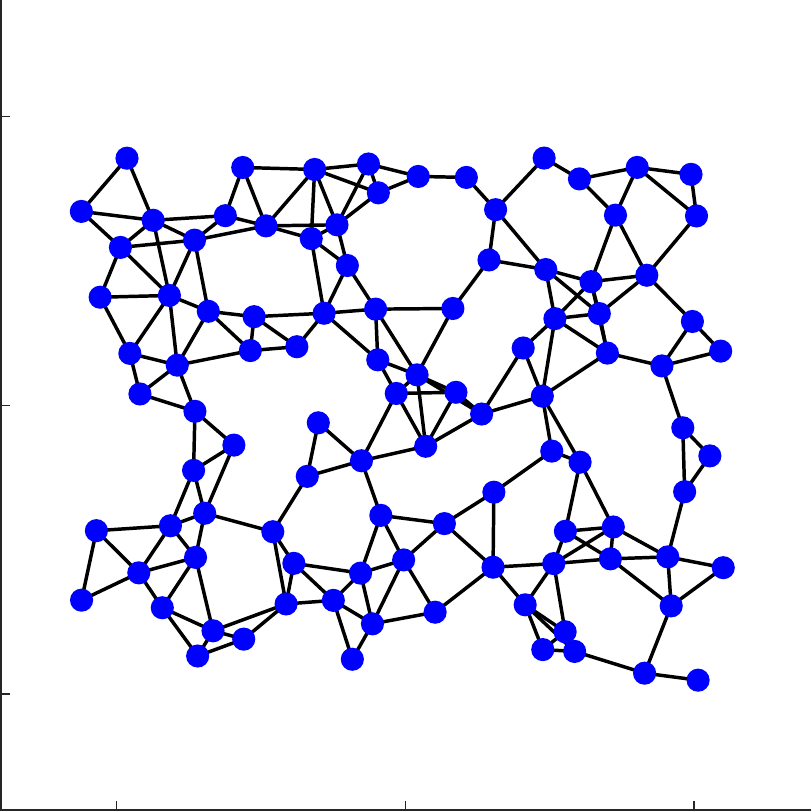}}
    \subfloat[$\delta=0.6$]{\includegraphics[trim={2mm 2mm 2mm 12mm},clip,width=0.32\columnwidth]{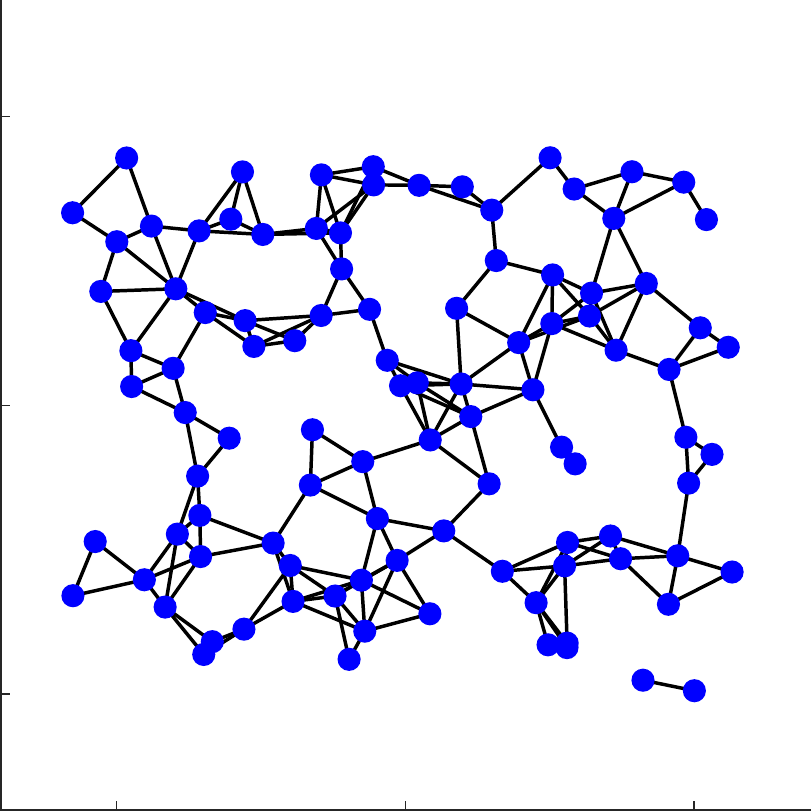}}
    \setcounter{subfigure}{2}}
    \\[-0.5ex]
    \subfloat[Final configurations]{
    \captionsetup[subfigure]{labelformat=empty}
    \subfloat[$\delta=0.2$]{\includegraphics[trim={2mm 2mm 2mm 12mm},clip,width=0.32\columnwidth]{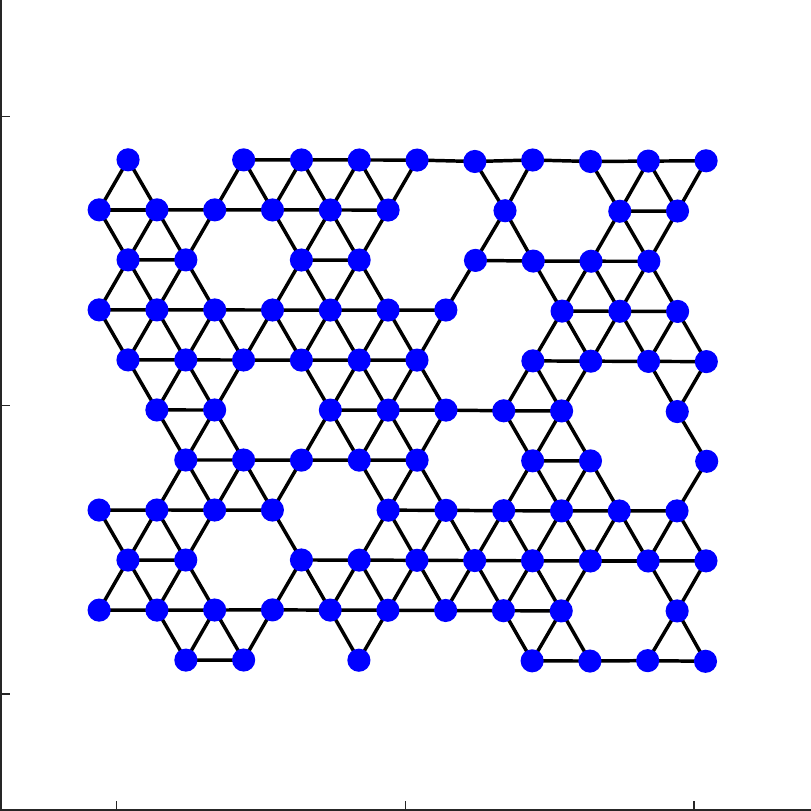}}
    \subfloat[$\delta=0.4$]{\includegraphics[trim={2mm 2mm 2mm 12mm},clip,width=0.32\columnwidth]{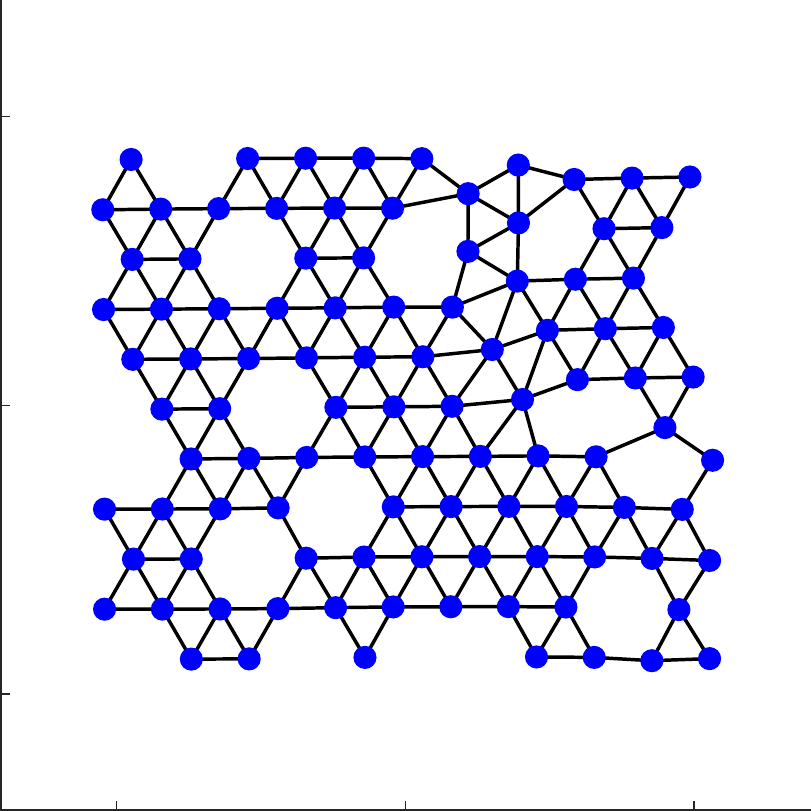}}
    \subfloat[$\delta=0.6$]{\includegraphics[trim={2mm 2mm 2mm 12mm},clip,width=0.32\columnwidth]{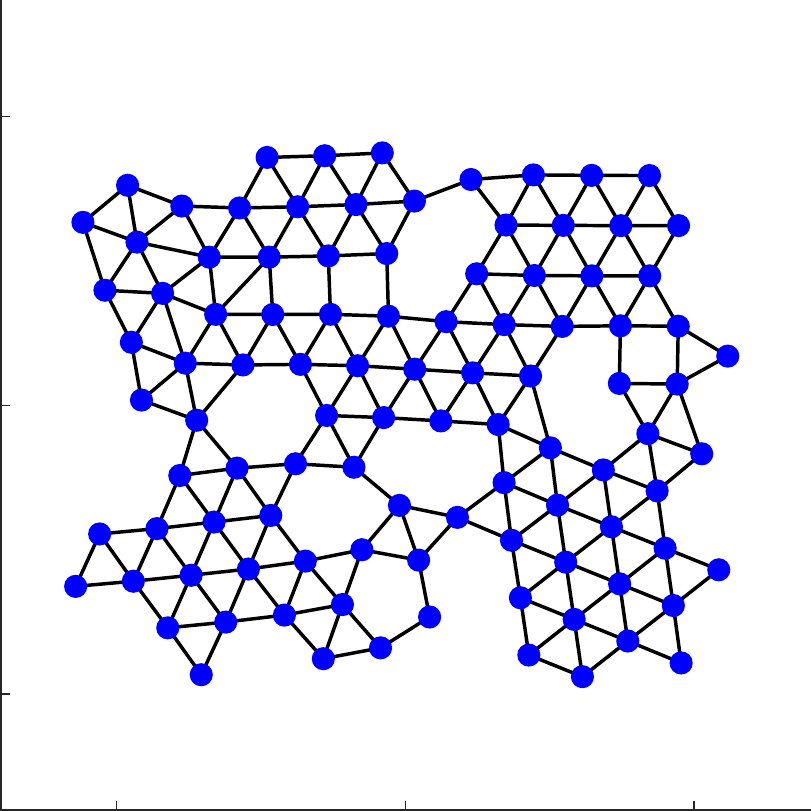}}
    \setcounter{subfigure}{3}}
    %
    \caption{Simulations for different values of $\delta$.
    (a): Terminal values of $e$ and $\rho$. 
    $\rho$ is the fraction of trials converging to an infinitesimally rigid configuration.
    For $e$, the solid line is the mean; the shaded area is the minimum and maximum.
    20 simulations with random initial conditions are performed for each value of $\delta$.
    (b), (c): Initial and final configurations of representative simulations for specific values of $\delta$.}
    \label{fig:varyingdelta}
\end{figure}

To validate Theorem \ref{th:local_stability_lyap} and estimate the basin of attraction of the set of triangular configurations, we performed extensive simulations for various values of $\delta$, and observed the steady state configurations. The results are reported in Fig.~\ref{fig:varyingdelta}.
Namely, we see that for $\delta \leq \delta^{\R{thres}} \coloneqq 0.25$ all simulations converge to a triangular configuration, with a rigid framework and a negligible value of $e$.
Then, as $\delta$ increases beyond $\delta^{\R{thres}}$, the average number of simulations converging to triangular configurations decreases, until for $\delta> 0.45$ no simulation converges to a triangular configuration.
Notice that $e(0)\leq 2\delta$, therefore $\delta = 0.25$ corresponds to a perturbation of up to 50\% of the initial length of the links, providing an estimation of the basin of attraction (region of asymptotic stability) of $\C{T}$.
%
%

\begin{figure}[t]
    \centering
    
    \includegraphics[width=0.9\columnwidth]{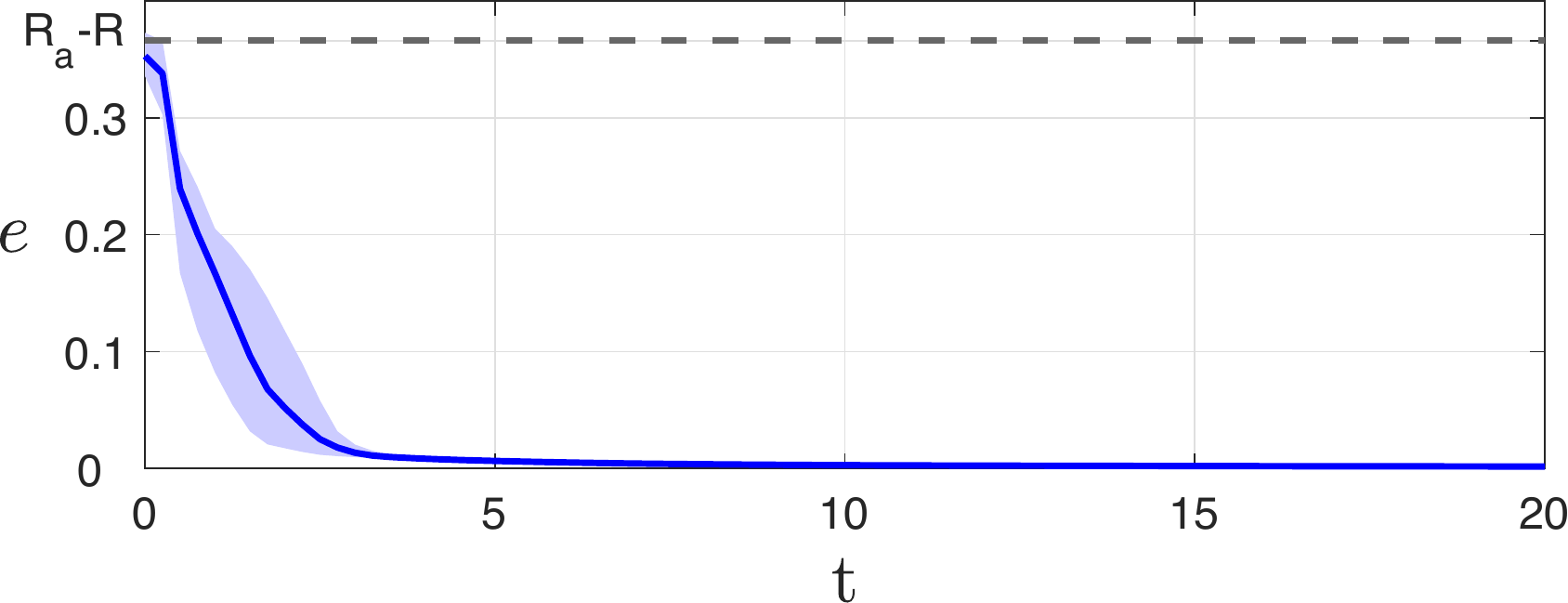}
    \caption{%
    Time evolution of the error $e$ in 10 simulations with random initial conditions and $\delta = 0.2$; the solid line is the mean, while the shaded area is the maximum and minimum.
    }
    \label{fig:simulation}
\end{figure}

Moreover, we analysed the time evolution of $e(t)$ in the case $\delta = 0.2$.
The results of $10$ simulations are shown in Fig.~\ref{fig:simulation}.
We find that the rigidity is preserved during all simulations, and at steady state $e$ reaches zero, meaning that the swarm, when locally perturbed, quickly converges back to a triangular configuration, as expected from Theorem~\ref{th:local_stability_lyap}.

\section{Conclusions}

We proved analytically local asymptotic stability of triangular lattice configurations for planar swarms under the action of a distributed control action based on virtual attraction/repulsion forces.
The theoretical derivations were supported by exhaustive numerical simulations validating the theoretical results and providing an estimate of the basin of attraction.
The mild hypotheses required on the interaction function that were used to prove convergence allow for wide applicability of the theoretical results.

Future work will focus on the formalization of the three-dimensional case and the extension of the results to other geometric lattices, such as squares and hexagons.





\appendix
\label{sec:appendix}

To confirm the effectiveness of our theoretical results, we provide below further semi-analytical evidence that the set of triangular configurations $\C{T}$, is locally asymptotically stable, which also excludes the presence of other equilibria in an arbitrarily small neighborhood of it.
To do so, we linearize
system \eqref{eq:model}--\eqref{eq:control_law} around a triangular configuration, say $\bar{\vec{x}}^*$, obtaining
$
    \label{eq:linearization}
    \dot{\bar{\vec{x}}} \approx \vec{J}(\bar{\vec{x}}^*) \, (\bar{\vec{x}}-\bar{\vec{x}}^*)
$, 
with $\vec{J}(\bar{\vec{x}}^*) \in \BB{R}^{2n \times 2n}$ derived as follows.

\paragraph*{Jacobian of \eqref{eq:model}--\eqref{eq:control_law}}
System \eqref{eq:model}--\eqref{eq:control_law} can be recast as
\begin{equation}\label{eq:model_stack}
    \dot{\bar{\vec{x}}} = ((\B{B} \vec{F} \B{G}^{-1} \B{B}\T) \otimes \B{I}_2) \bar{\vec{x}}= ((\B{B} \vec{H} \B{B}\T) \otimes \B{I}_2) \bar{\vec{x}},
\end{equation}
where $\B{F}, \B{G}, \B{H} \in \BB{R}^{m \times m}$ are diagonal matrices; $[\B{F}]_{ii} \coloneqq f(\norm{\vec{r}_i})$, $[\B{G}]_{ii} \coloneqq \norm{\vec{r}_i}$, and $\vec{H} \coloneqq \B{F} \B{G}^{-1}$. 
The Jacobian of \eqref{eq:model_stack} is 
\begin{equation}
\label{eq:Jacobian}
\begin{aligned}    
    \vec{J}&=\left(
    \vec{B} \frac{\partial \vec{H}}{\partial \bar{\vec{x}}}  \vec{B}\T \otimes \vec{I}_2 \right) \vec{\bar{x}} +
    (\B{B} \vec{H} \B{B}\T) \otimes \B{I}_2 \eqqcolon \vec{J}_1 +\vec{J}_2,
\end{aligned}
\end{equation}
where  $\frac{\partial \vec{H}}{\partial \bar{\vec{x}}} \in \BB{R}^{m\times m \times 2n}$ is a tensor, and 
\begin{equation*}
    \left[\frac{\partial \vec{H}}{\partial \bar{\vec{x}}}  \B{B}\T \right]_{:,:,k} =  \left[ \frac{\partial \vec{H}}{\partial \bar{\vec{x}}} \right]_{:,:,k} \B{B}\T \quad \in \BB{R}^{m\times n},
\end{equation*}
with notation $[\ \cdot\ ]_{:,:,k}$ denoting the matrix obtained by fixing the third index of the tensor.
From \ref{hp:null_point}, for all triangular configurations we have $\vec{J}_2=(\B{B} \vec{H} \B{B}\T) \otimes \B{I}_2 =\vec{0}$.
Then,
$
    \left[ \vec{J}_1 \right]_{:,k} 
    = \left(
    \vec{B} \left[ \frac{\partial \vec{H}}{\partial \bar{\vec{x}}} \right]_{:,:,k}  \vec{B}\T \otimes \vec{I}_2 \right) \vec{\bar{x}}
$. 
From {\cite[p. 20]{Jackson2007} 
}%
we have 
$\frac{\partial \norm{\vec{r}_i}^2}{\partial [\bar{\vec{x}}]_k}  = 2 [\vec{M}]_{i,k}$ (see Definition \ref{def:rigidity_matrix}),
that is
$\frac{\partial \norm{\vec{r}_i}}{\partial [\bar{\vec{x}}]_k}  = \frac{1}{\norm{\vec{r}_i}} [\vec{M}]_{i,k}$, and thus
\begin{subequations}\label{eq:matrix_H}
\begin{align}
    \left[ \frac{\partial \vec{H}}{\partial \bar{\vec{x}}} \right]_{i,i,k} 
    &= \frac{\partial [f(\norm{\vec{r}_i})/\norm{\vec{r}_i}]}{\partial \norm{\vec{r}_i}} \frac{\partial \norm{\vec{r}_i}}{\partial [\bar{\vec{x}}]_k}\nonumber\\
    &= [f'(\norm{\vec{r}_i}) \norm{\vec{r}_i}  - f(\norm{\vec{r}_i})] \norm{\vec{r}_i}^{-3} [\vec{M}]_{i,k},\\
    \left[ \frac{\partial \vec{H}}{\partial \bar{\vec{x}}} \right]_{i,j,k} &= 0, \quad \text{if} \ i \ne j.
\end{align}
\end{subequations}

\paragraph*{Numerical analysis}
We set $R = 1$ and generated $760$ random triangular configurations ($10$ per each number of agents 
$n$ between $25$ and $100$).
For each of these configurations, assuming $f$ (in \eqref{eq:control_law}) is in the form \eqref{eq::Lennard-Jones}, we computed $\B{J}$ using \eqref{eq:Jacobian}--\eqref{eq:matrix_H} and found that in all cases $\vec{J}$ has $3$ zero eigenvalues with eigenvectors $\{\vec{w}_i^{0}\}_{i}$, and $2n-3$ negative eigenvalues with eigenvectors $\{\vec{w}_j^{\pm}\}_j$.
Moreover, $\B{M} \B{w}_i^{0} = \B{0}$ and $\B{M} \B{w}_j^{\pm} \ne \B{0}$; thus, from Definition~\ref{def:inf_rigidity}, the span of $\{\B{w}_i^{0}\}$ corresponds to roto-translations and is a hyperplane locally tangent to $\Gamma(\bar{\vec{x}}^*)$ (see Definition \ref{def:congruent_conf}), while $\{\B{w}_j^{\pm}\}$ correspond to other motions.
Therefore, the \emph{center manifold theorem} \cite[Theorem 5.1]{Kuznetsov2004} yields that $\Gamma(\bar{\vec{x}}^*)$ is a \emph{center manifold} of system \eqref{eq:model}--\eqref{eq:control_law}.
Moreover, as expected from Theorem \ref{th:local_stability_lyap}, the \emph{reduction principle} \cite[Theorem 5.2]{Kuznetsov2004} confirms that the dynamics locally converge onto the 
equilibrium set $\Gamma(\bar{\vec{x}}^*)$, and excludes the presence of other equilibria in an  arbitrarily small neighborhood of it.




\bibliographystyle{IEEEtran}
\bibliography{LSMAS, Synt_Bio}


\end{document}